\documentclass{cimento}

\usepackage{graphicx}  % got figures? uncomment this

\title{Testing Predictions of the Chiral Anomaly in Primakoff Reactions at COMPASS}
\author{D.~Ecker\thanks{dominik.ecker@tum.de} \textup{for the COMPASS collaboration}}
\instlist{\inst{} Technical University of Munich - Munich, Germany}

\def\slh#1{#1\kern-.65em\slash}

\begin{document}

\maketitle

\begin{abstract}
  The chiral anomaly is a fundamental property of quantum chromodynamics (QCD). It governs the transition amplitudes for processes involving an odd number of Goldstone bosons of chiral symmetry breaking. In case of the coupling of three pions to a photon, the magnitude of the resulting coupling is $F_{3\pi}$ and the value is predicted by chiral perturbation theory with small uncertainty. It can experimentally be measured in $\pi^-\gamma \to \pi^- \pi^0$ scattering.

  Here, we report on a precision experiment on $F_{3\pi}$ using the COMPASS experiment at CERN where pion-photon scattering is mediated via the Primakoff effect using heavy nuclei as target. We exploit the interference of the production of the $\pi^- \pi^0$ final state via the chiral anomaly with the photo-production of the $\rho(770)$ resonance over a wide mass range $M_{\pi^- \pi^0}<1\textrm{ GeV}/c^2$. This is in contrast to previous measurements restricting themselves only to the threshold region $M_{\pi^- \pi^0}<370\textrm{ MeV}$. Our analysis allows to simultaneously extract the radiative width of the $\rho(770)$ resonance and gives a stronger handle on $F_{3\pi}$ in a unified approach thereby minimizing systematic effects rarely addressed previously.
\end{abstract}

\section{The chiral anomaly}

%The theory of strong interaction, quantum chromodynamics (QCD), is built to exhibit local color symmetry. % representing the fact, that strong interaction couples equally to red, green, and blue quarks.
% The Lagrange density of the theory of strong interaction, quantum chromodynamics (QCD), is given by
% \begin{equation}
%   \mathcal{L}_{QCD} = \sum_{f} \bar{q}_f \left( i \gamma^\mu D_\mu - m_f \right)q_f - \frac{1}{2} \Tr \left(G_{\mu\nu} G^{\mu\nu}\right)
%   \label{eq:L_QCD}
% \end{equation}
% with the quark spinor $q_f$, the covariant derivative $D_\mu = \partial_\mu + i g_s A_\mu$, $\Tr\left( G_{\mu\nu}G^{\mu\nu} \right) = \frac{1}{2} G_{\mu\nu}^a G_a^{\mu\nu}$, $G_a^{\mu\nu} = \partial^\mu A_a^\nu - \partial^\nu A_a^\mu + g f_{abc} A_b^\mu A_c^\nu$, and the gluon field $A_\mu = \sum_{a=1}^8 \frac{\lambda_a}{2} A_\mu^a$. Since all flavors have (slightly) different masses, the mass term breaks flavor symmetry explicitely.

The difference in masses $m_f$ in the quark fields of the Lagrangian of quantum chromodynamics (QCD)
\begin{equation}
  \mathcal{L}_{QCD} = %\sum_{f} 
  \bar{q}_f \left( i \gamma^\mu \left( \partial_\mu + ig A_\mu \right) - m_f \right)q^f
  \label{eq:L_QCD}
\end{equation}
explicitly breaks flavor symmetry. % for different $m_f$.
Exploiting the smallness of the quark masses for the flavors $f=u,d,s$ with respect to a typical hadronic scale, and exploring QCD in the limit of vanishing quark masses $m_f \to 0$, the so-called \textit{chiral limit}, shows that left- and right-handed quark fields, $q_L$ and $q_R$, decouple
$\mathcal{L}_{QCD} = \mathcal{L}_{QCD}(q_L) + \mathcal{L}_{QCD}(q_R)$.
We observe a flavor symmetry of $u$, $d$, and $s$ flavor for $q_L$ and $q_R$:
\begin{equation}
  SU(3)_L \times SU(3)_R.
  \label{eq:chiralSymmetry}
\end{equation}

The spontaneous breaking of chiral symmetry gives rise to the octet of light pseudoscalar mesons $\left( \pi^\pm, \pi^0, K^\pm, K^0, \bar{K}^0, \eta \right)$, which appear to be unnaturally light. 
%Indeed, if chiral symmetry was exact, \textit{i.e.} $m_f = 0$, also the Goldstone bosons of the symmetry breaking would be massless. However, chiral symmetry is only approximate and explicitly broken by the small but nonzero quark masses. Hence, also the pseudoscalar Goldstone bosons acquire mass.

%The most fundamental conservation laws arise as a consequence of symmetries of the underlying Lagrangian. A symmetry can also be spontaneously broken. Then, it must give rise to corresponding Goldstone bosons. 
The chiral flavor symmetry of eq.~(\ref{eq:chiralSymmetry}) is a good example of how symmetries of the classical Lagrangian manifest themselves, albeit only approximately, in the quantum world. However, there are symmetries of the classical Lagrangian that are not respected in the quantum realm, so-called anomalies. The most famous is the chiral anomaly, which is associated with the axial $U(1)$ symmetry of QCD in the chiral limit. Despite the symmetry, the axial current is not conserved in the chiral limit.
%: transformations of the form
%\begin{equation}
%  q(x) \to e^{i\theta\gamma_5} q(x)
%\end{equation}
%leave the Lagrangian invariant.

The chiral anomaly drives processes with an odd number of pseudoscalars (or axial currents), such as the $\pi^0$ decay~\cite{ref:Bernstein}, for which the anomalous prediction has been confirmed by various experiments~\cite{ref:PDG}.
% In fact, the chiral anomaly was discovered when scientists were trying to explain the-at the time-astonishing short $\pi^0$-lifetime of \cite{ref:PDG}
% \begin{equation}
%   \tau_\mathrm{PDG}(\pi^0) = \left( 8.52 \pm 0.18 \right) \cdot 10^{-17} \mathrm{s}.
%   \label{eq:tau_PDG}
% \end{equation}
% The anomaly increases the coupling of processes involving an odd number of pseudoscalars, such as the $\pi^0$-decay and its prediction
% \begin{equation}
%   \tau_\mathrm{anom}(\pi^0) = 8.38 \cdot 10^{-17} \rm{s}
% \end{equation}
% matches spectacularly well with the averaged experimental value of eq.~(\ref{eq:tau_PDG}).

\begin{figure}
  \centering
  \includegraphics[scale=0.8]{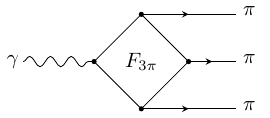}
  % \hfill
  \hspace{1cm}
  \includegraphics[scale=0.8]{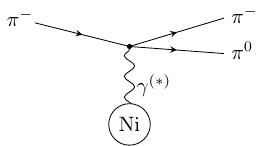}
  \caption{On the left: tree-level diagram of the anomalous coupling of one photon to three pions. The coupling strength is described by the chiral anomaly constant $F_{3\pi}$. On the right: Primakoff reaction, in which an ultra-relativistic beam pion scatters in the Coulomb field of a nucleus giving rise to the photon, three pion vertex.}
  \label{fig:Feynman}
\end{figure}

Other anomalous processes lack similarly precise experimental verification. % despite the chiral anomaly being such a fundamental property of QCD.
The next most accessible anomalous coupling is the direct coupling of three pions to one photon as shown in Fig.~\ref{fig:Feynman}. 
The anomalous prediction for the coupling strength of this process
\begin{equation}
  F_{3\pi}^\mathrm{theory} = \frac{e N_c}{12 \pi^2 F_\pi^3} = \left( 9.78 \pm 0.04 \right) \rm{GeV}^{-3}
  \label{eq:F3pi_theory}
\end{equation}
depends on only three parameters, the elementary charge $e$, the number of colors $N_c$, and the pion decay constant $F_\pi = \left( 92.21 \pm 0.14 \right)\rm{MeV}$. 
The coupling can be accessed experimentally through
\begin{equation}
  \pi^- \gamma \to \pi^- \pi^0
  \label{eq:reaction}
\end{equation}
reactions. 
The COMPASS collaboration has studied pion-photon interactions and performed an up-to-date measurement of $F_{3\pi}$.
%There exists a measurement dating back to the 1980ies by Antipov \textit{et al.}~\cite{ref:Antipov}, which first suggested tension with the low-energy theorem of eq.~(\ref{eq:F3pi_theory}).
%\begin{equation}
%  F_{3\pi}^\mathrm{Antipov}  = \left( 12.9 \pm 0.9 \pm 0.5 \right) \rm{GeV}^{-3}
%\end{equation}

\section{COMPASS measurement}

For this measurement, the COMPASS experiment at CERN uses secondary hadron beams provided by the super proton synchrotron (SPS). The multi-purpose detector setup is designed to support a wide range of research in hadron physics. 
It provides high-precision and high-rate tracking capabilities, efficient particle identification, and precise calorimetry.
% in both stages of the spectrometer, each centered around a dipole magnet. The two stages are purposefully optimized for low and high momentum particles, enabling precise momentum determination with an accuracy of better than 1\% across a wide range, spanning from approximately 1 GeV up to the beam's momentum range, which extends to 200 GeV.

\subsection{Primakoff effect}

The method used to study pion-photon interactions exploits the strong electric field in the vicinity of atomic nuclei. The field acts as a source of quasi-real photons, when viewed from an ultra-relativistic reference frame.
Initially, the idea was conceived by H. Primakoff~\cite{ref:Primakoff} to measure the $\pi^0$ lifetime through photon-photon fusion. It was later realized that interactions between high-energy hadrons and the nuclear Coulomb field similarly represent the exchange of a single photon between nucleus and beam hadron. 

% The method employed in investigating pion-photon interactions is credited to H. Primakoff~\cite{ref:Primakoff}. 
% It exploits the strong electric field in the vicinity of atomic nuclei, which acts as a source of quasi-real photons, when viewed from an ultra-relativistic reference frame.
% Initially, the idea was conceived for measuring the lifetime of the $\pi^0$ through photon-photon fusion. It was later realized that interactions between high-energy hadrons and the nuclear Coulomb field similarly represent the exchange of a single photon between nucleus and beam hadron. 
%Extending the original idea, one considers Primakoff reactions or the Primakoff effect as scattering of any ultra-relativistic particle, not only photons, on Coulomb-field quanta, i.e. quasi-real photons, of a nucleus.

At COMPASS, we use nickel as target material, which provides sufficiently high $Z$ for a strong Coulomb field and negligible radiative corrections due to double photon exchange or screening.

Primakoff reactions are characterized by low-momentum transfer $Q = \sqrt{Q^2} = \sqrt{-q^\mu q_\mu}$ with $q^\mu$ being the four-momentum of the exchanged photon. 
At COMPASS beam energies of $p_\mathrm{beam} = 190 \rm{GeV}/c$, typical values for $Q$ are in the order of few MeV. The scale difference of five orders of magnitude to the beam momentum requires precise calibration of the spectrometer.

\subsection{Resonant production of the $\rho(770)$ meson}

In addition to the dominant coupling to two pions, the $\rho(770)$ also couples to $\pi\gamma$ with a small branching fraction. % of $\Gamma_{\pi\gamma}/\Gamma_{\mathrm{tot}} = \left( 4.5 \pm 0.5 \right) \cdot 10^{-4}$~\cite{ref:PDG}. 
It can therefore be resonantly produced in the $s$-channel in $\pi^-\gamma \to \rho^- \to \pi^-\pi^0$ events constituting a fully coherent background to the direct coupling of a photon to three pions. The $\sqrt{s}$-dependence of the cross section for $\pi^-\gamma \to \pi^-\pi^0$ is thus a result of the interference between resonant $\rho$ production and chiral production of the $\pi^-\pi^0$ final state. 
%as depicted in fig.~\ref{fig:rho}.

% \begin{figure}
%   \centering
%   % \includegraphics{Images/BG_rhoInterference.pdf}
%   \includegraphics[width=0.5 \textwidth]{Images/Kaiser}
%   \caption{Total cross sections for the reaction $\pi^-\gamma \to \pi^-\pi^0$ as a function of the center-of-mass energy $\sqrt{s}$ calculated by Kaiser and Friedrich in~\cite{ref:Kaiser}. The blue line shows the prediction of the anomalous chiral production on tree-level. The red line includes additionally a model for the $\rho$-resonance.}
%   \label{fig:rho}
% \end{figure}

The analysis presented uses the dispersive approach worked out in Ref.~\cite{ref:Kubis}. The framework extends the useful kinematic range from the chiral low-mass region up to $\sim 1~\rm{GeV}$ beyond the $\rho$ resonance. Additional information hidden in the shape of the resonance peak due to the interference with the chiral production can be exploited in this way.

\subsection{Extrapolation in chiral perturbation theory and radiative corrections}

The value of eq.~(\ref{eq:F3pi_theory}) is defined in the chiral limit at $s=t=u=0$. Scattering experiments according to eq.~(\ref{eq:reaction}) can only access the region $s>4m_\pi^2$. One- and two-loop corrections in chiral perturbation theory, which became available in~\cite{ref:Hannah}, introduce an $s$ dependence and are necessary to extrapolate from the kinematic range covered by experiment to the chiral limit. Finally, we also have to consider the large radiative corrections which are mainly due to $t$-channel exchange of a photon~\cite{ref:Knecht}.

\subsection{Background subtraction}

Potential background arises from production of the $\pi^-\pi^0$ final state via the strong interaction.
Direct production of the $\pi^-\pi^0$ final state via Pomeron exchange is not possible due to $G$-parity conservation. The cross section of the exchange of ordinary Reggeons is small at COMPASS beam energies.
%One of the challenges of the presented analysis is to control the background to the $\pi^-\gamma \to \pi^-\pi^0$ channel. 
The largest background contribution stems from $\pi^- \rm{Ni} \to \pi^-\pi^0\pi^0\rm{Ni}$ reactions. They can proceed via Pomeron exchange and lead to a large cross section. Even a small probability that one of the $\pi^0$s is low-energetic and remains undetected, can become significant in the Primakoff region of small $Q$.
We exploit the different shape of the background in $Q^2$ to disentangle $3\pi$ background from signal. 
To obtain a realistic model for the leakage, we perform a full analysis of the $\pi^- \rm{Ni} \to \pi^-\pi^0\pi^0\rm{Ni}$ reaction fitting a partial-wave model to the data.
This allows us to get a model, which is differential in all kinematic quantities, and which can be used to extrapolate the number of $\pi^-\pi^0\pi^0$ events into the kinematic range, in which one of the $\pi^0$s is too low-energetic to be detected. 
%We obtain a prediction for the leakage from this corner of the phase space.

\begin{figure}
  \centering
  \includegraphics[width=0.93\textwidth]{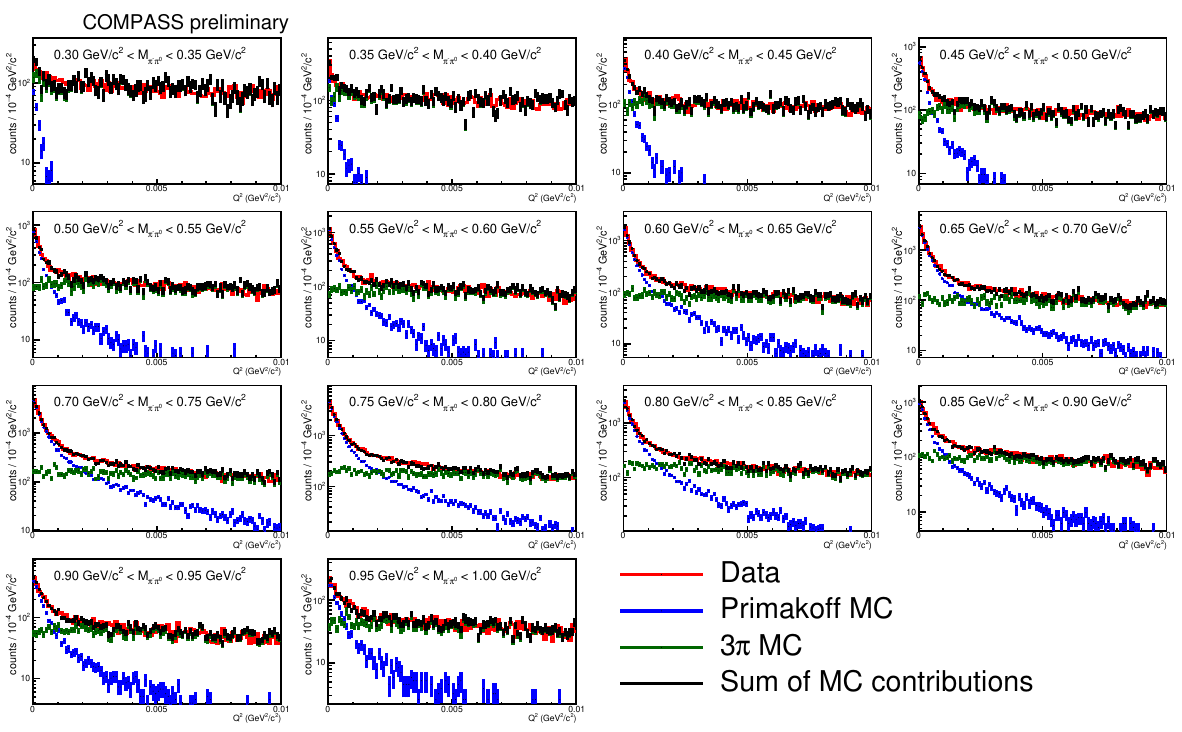}
  \caption{Measured $Q^2$-distribution in different mass bins of the two-body final state and comparison to Monte-Carlo simulation consisting of two components: Primakoff signal and three-pion background.}
  \label{fig:bg}
\end{figure}

The background lacks the characteristic Primakoff peak at very low $Q^2$, and becomes dominant at larger $Q^2$. Fitting the observed $Q^2$ distributions with simulated distributions for signal as well as for $3\pi$-leakage in bins of $M_{\pi^-\pi^0}$, as it is shown in fig.~\ref{fig:bg}, yields the number of $\pi^-\gamma \to \pi^-\pi^0$ events in each bin. 
A consistent description of the data could be achieved with only these two components. 
Currently, we are investigation additional background components to further improve our analysis results.
%We are still investigating additional background components, so this part of the analysis is still ongoing.

\subsection{Determination of the luminosity}

We determined the luminosity from decays of free kaons, which occur in the secondary negative hadron beam with a known fraction of $\left( 2.79 \pm 0.05 \right)\%$. 
Using the free kaon decays, we determine an effective luminosity, which naturally takes into account the effects of trigger and DAQ deadtimes. %, in the same way as beam pions.
In addition, $K^- \to \pi^- \pi^0$ decays show similar systematic effects as the $\pi^-\gamma \to \pi^- \pi^0$ reaction, since both processes share the same final state.
$K^- \to \pi^-\pi^0\pi^0$ decays are used to check for systematic uncertainties.
The preliminary result for the integrated luminosity
\begin{equation}
  L_\mathrm{eff} = \left( 5.21 \pm 0.04_\mathrm{stat} \pm 0.48_{\mathrm{syst}} \right)~\rm{nb}^{-1}
\end{equation}
is dominated by systematic uncertainties. These can be reduced by further analysis.

\subsection{Preliminary result}

\begin{figure}
  \centering
  \includegraphics[width=0.55\textwidth]{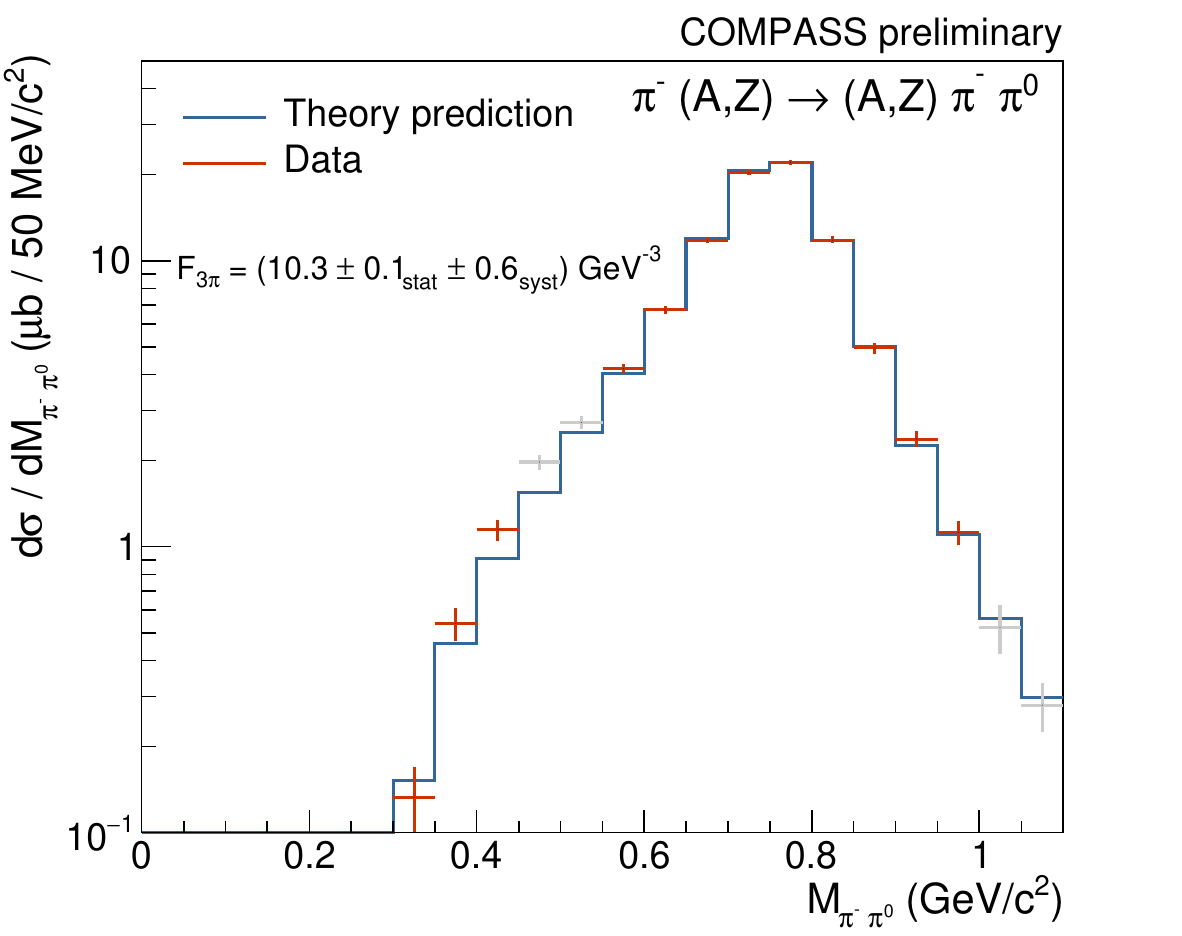}
  \caption{The preliminary COMPASS result for the cross section of $\pi\gamma \to \pi\pi$ in bins of the invariant mass of the final state. The data is indicated in red, the prediction of the dispersive framework for each bin is drawn as blue line. The region around $M_{\pi^-\pi^0} = 0.5 \rm{GeV}/c^2$ affected by the decays of beam kaons $K^-\to\pi^-\pi^0$ is excluded in the fit and the corresponding data points are shown in grey.}
  \label{fig:result}
\end{figure}

With the described preliminary background subtraction, we obtain the $M_{\pi^-\pi^0}$ distribution shown in red in fig.~\ref{fig:result}. By using the dispersive model of~\cite{ref:Kubis}, we obtain 
\begin{equation}
  F_{3\pi}^\mathrm{preliminary} = \left( 10.3 \pm 0.1_\mathrm{stat} \pm 0.6_\mathrm{syst}\right) \rm{GeV}^{-3}.
\end{equation}
The result is in agreement with the anomaly prediction of eq.~(\ref{eq:F3pi_theory}). The result has large systematic uncertainties, mainly due to the luminosity determination and the background subtraction.
Still, the uncertainty is unprecedented and a factor two smaller than in the previous determinations of Refs.~\cite{ref:Knecht,ref:Moinester}, which are based on reanalysis of Refs.~\cite{ref:Antipov,ref:Amendolia} and do not provide an estimate on systematic uncertainties at all.
%  do not provide an estimate of the systematic uncertainties. 
%The most recent reanalysis of the previous data in Ref.~\cite{ref:Knecht} does not give any estimate on the systematics. % , despite being probably dominated by systematic uncertainties.
We still aim to reduce the systematic uncertainty significantly. 

Fitting the dispersive model also allows to obtain the radiative width of the $\rho$~\cite{ref:Hoferichter}
\begin{equation}
  \Gamma_{\rho\to\pi\gamma}^\mathrm{preliminary} = \left( 
    76 {\pm 1_\mathrm{stat}} \left(^{+10}_{-8}\right)_\mathrm{syst} 
  \right)~\rm{keV}.
\end{equation}
The value obtained is higher than the PDG average of $\left(67 \pm 8 \right)~\rm{keV}$ consisting of three previous measurements~\cite{ref:Capraro,ref:Huston,ref:Jensen}. However, these measurements differ on a $\sim 25\%$-level. The COMPASS result is lower than the most recent one, which used a similar experimental method~\cite{ref:Capraro}, but neglected the chiral production of the $\pi^-\pi^0$ final state.

%\section{Examples}

% \begin{table}
  % \caption{Prices of important items.}
  % \label{tab:pricesI}
  % \begin{tabular}{rcl}
  %   \hline
  %     Item 1      & 1500  & EUR    \\
  %     Item 2 & 15000 & EUR    \\
  %     Item 3      & 1500  & dollars \\
  %   \hline
  %     Item 4     & .25   & dollars \\
  %     Item 5         & 1.25  & dollars \\
  %     Item 6         & 1     & dollars \\
  %   \hline
  % \end{tabular}
% \end{table}

% \subsection{Mathematics}
% Here is a lettered array~(\ref{e.all}), with eqs.~(\ref{e.house})
% and~(\ref{e.phi}):
% \begin{eqnletter}
%  \label{e.all}
%  \drm x_\sy{F} & = & 1.2\cdot10^3\un{cm}, \qquad
%                      \tx{where\ } \sy{F} = \tx{Fermi}    \label{e.house}\\
%  \phi_i        & = & i\pi                                \label{e.phi}
% \end{eqnletter}

\acknowledgments
I would like to thank Andrii Maltsev, Dmitry Ryabchikov, and Jan Friedrich who contributed substantially. The research was funded by the DFG under Germany’s Excellence Strategy EXC2094-390783311 and BMBF Verbundforschung 05P21WOCC1 COMPASS.

\end{document}